\documentclass[aps,prb,floatfix,epsfig,twocolumn,showpacs,preprintnumbers,superscriptaddress]{revtex4}
\usepackage{amsmath}
\usepackage{graphicx}
\usepackage{dcolumn}
\usepackage{bm}

\input{tcilatex}
\begin{document}

\title{Electronic structure and magnetic properties of NaOsO$_{3}$}

\author{Yongping Du}
\affiliation{National Laboratory of Solid State Microstructures and Department of
Physics, Nanjing University, Nanjing 210093, China}

\author{ Xiangang Wan}
\thanks{Corresponding author: xgwan@nju.edu.cn}
\affiliation{National Laboratory of Solid State Microstructures and Department of
Physics, Nanjing University, Nanjing 210093, China}
\affiliation{Department of Physics, University of California, Davis, One Shields
Avenue, Davis, CA 95616.}

\author{Li Sheng}
\affiliation{National Laboratory of Solid State Microstructures and Department of
Physics, Nanjing University, Nanjing 210093, China}

\author{Jinming Dong}
\affiliation{National Laboratory of Solid State Microstructures and Department of
Physics, Nanjing University, Nanjing 210093, China}

\author{ Sergey Y. Savrasov}
\affiliation{Department of Physics, University of California, Davis, One Shields
Avenue, Davis, CA 95616.}

\begin{abstract}
A comprehensive investigation of the electronic and magnetic properties of
NaOsO$_{3}$ has been made using the first principle calculations, in order
to understand the importance of Coulomb interaction, spin--orbit coupling
and magnetic order in its temperature--induced and magnetic--related
metal--insulator transition. It is found that its electronic structure near
the Fermi energy is dominated by strongly hybridized Os 5\textit{d} and O\ 2%
\textit{p} states. Despite of the large strength of spin--orbit coupling, it
has only small effect on the electronic and magnetic properties of NaOsO$_{3}
$. On the other hand, the on--site Coulomb repulsion affects the band
structure significantly, but, a reasonable \textit{U} alone cannot open a
band gap. Its magnetism is itinerant, and the magnetic configuration plays
an important role in determining the electronic structure. Its ground state
is of a G--type antiferromagnet, and it is the combined effect of \textit{U}%
\ and magnetic configuration that results in the insulating behavior of NaOsO%
$_{3}$.
\end{abstract}

\date{\today }
\maketitle

\section {INTRODUCTION}
It is well known that the Coulomb interaction among 3\textit{d} electrons in
transition--metal oxides (TMO) is substantially important, which induces
peculiar properties, such as metal--insulator transition\cite{Review M-I},
colossal magnetoresistance\cite{CMR} and high critical temperature
superconductivity\cite{Pickett-HTC}. Because the 5\textit{d} orbitals are
highly extended compared to those in the 3\textit{d} systems, it is natural
to expect that the electronic correlations are weak and have only negligible
effect in the 5\textit{d} compounds. However, recent theoretical and
experimental works have given the evidence on the importance of Coulomb
interactions here\cite{Sr2IrO4-1,Sr2IrO4-2}. On the other hand, the
spin--orbit coupling (SOC) in the 5$d$ transition metal elements is expected
to be strong due to the large atomic number\cite{SOC}. Hence, due to the
interplay of electron correlations and strong spin--orbit interactions,
various anomalous electronic properties have been observed/proposed in the 5%
\textit{d} transition oxides, such as J$_{eff}$=1/2 Mott state\cite%
{Sr2IrO4-1,Sr2IrO4-2}, giant magnetoelectric effect\cite{GME}, high T$_{c}$
superconductivity\cite{HTC}, Weyl semimetal with Fermi arcs\cite%
{Weyl-semimetal}, Axion insulator with large magnetoelectric coupling\cite%
{Axion insulator},\ topological insulator\cite{Pesin-Balents,Kim},
correlated metal\cite{Bi2Ir2O7}, Kitaev mode\cite{Kitav}, etc.

One class of the well studied 5\textit{d} compounds are the osmates \cite%
{Pickett,NaOsO3, Cd2Os2O7-1,Cd2Os2O7, Singh LSDA,RbOs2O6,BaNiOsO,KOs2O6
DFT,Cd2Os2O7-3,Cd2Os2O7-4,Cd2Os2O7-5}. For example, the physical properties
of Cd$_{2}$Os$_{2}$O$_{7}$ are quite intriguing. It has been found that Cd$%
_{2}$Os$_{2}$O$_{7}$ is metallic at room temperatures, while undergoing a
metal--insulator transition (MIT) at about 230 K\cite{Cd2Os2O7-1}.
Experiments reveal that this MIT is continuous and purely electronic.
Moreover, it is coincident with a magnetic\ transition of antiferromagnetic
(AFM)\ character\cite{Cd2Os2O7-1, Cd2Os2O7}. Therefore, experimentalists
argue that Cd$_{2}$Os$_{2}$O$_{7}$ is the first well--documented example of
a pure Slater transition\cite{Cd2Os2O7,Slater state}. However, despite of
the vast efforts devoted\cite{Cd2Os2O7-1, Cd2Os2O7, Cd2Os2O7-3,
Cd2Os2O7-4,Singh LSDA,Cd2Os2O7-5}, its exact magnetic ground state
configuration is still unknown due to the strong geometric frustration of
the pyrochlore lattice. Therefore, theoretical evidence of the Slater
transition in this compound is still lacking.

Recently, using high pressure technique, Shi \textit{et al.}\cite{NaOsO3}
synthesized another osmate: NaOsO$_{3}$. Similar to Cd$_{2}$Os$_{2}$O$_{7}$,
NaOsO$_{3}$ also exhibits a temperature--induced MIT, which is again
accompanied by a magnetic ordering without any lattice distortion\cite%
{NaOsO3}. However, better than Cd$_{2}$Os$_{2}$O$_{7}$, NaOsO$_{3}$ has a
simple perovskite structure, consequently being free from the complication
induced by magnetic frustration. Therefore, NaOsO$_{3}$\ provides a unique
platform to understand the temperature--induced and magnetic--related MIT.
Based on the experimental crystal structure, Shi \textit{et al.}\cite{NaOsO3}
also perform the band-structure calculation for this compound. They\cite%
{NaOsO3} find that both LDA\ and LDA+SO calculation give the paramagnetic
solution. Their numerical results\cite{NaOsO3} show that Coulomb U\ is not
efficient, and antiferromagnetic correlation is essential to open the band
gap\cite{NaOsO3}. Recently, there are experimental and theoretical evidences
of the importance of electronic correlation and spin-orbital coupling in 5d
transition-metal compounds. Therefore a comprehensive investigation of the
effect of the Coulomb interaction, SOC and magnetic order on its electronic
structure and MIT is still an interesting problem which we address in the
present work.

\section {METHOD}
The electronic band structure calculations have been carried out by using
the full potential linearized augmented plane wave\ method as implemented in
WIEN2K package\cite{WIEN}. Local spin density approximation (LSDA) is widely
used for various 4$d$ and 5$d$ transition metal oxides\cite%
{Sr2IrO4-1,Sr2IrO4-2,LSDA-good-1,LSDA-good-2,KOs2O6 DFT}, and we therefore
adopt it as the exchange--correlation potential. The muffin--tin radii for
Na, Os and O are set to 1.13, 1.02, and 0.90 \AA , respectively. The basic
functions are expanded to R$_{mt}$K$_{max}$=7 (where R$_{mt}$ is the
smallest of the muffin-tin sphere radii and K$_{max}$ is the largest
reciprocal lattice vector used in the plane wave expansion), corresponding
to 1915 LAPW functions at the $\Gamma $\ point. Using the second--order
variational procedure\cite{SOC in WIEN2k}, we include the spin--orbital
coupling interaction (SOC), which has been found to play an important role
in the 5\textit{d} system\cite{Sr2IrO4-1,Sr2IrO4-2,Weyl-semimetal,Axion
insulator}. A 10$\times $6$\times $10 mesh is used for the Brillouin zone
integral. The self--consistent calculations are considered to be converged
when the difference in the total energy of the crystal does not exceed 0.1
mRy and that in the total electronic charge does not exceed 10$^{-3}$
electronic charge at consecutive steps.

\begin{table}[tbph]
\caption{Numerical and experimental internal coordinates of NaOsO$_{3}$.}%
\centering%
\begin{tabular}{c c c c c c c c}
\hline
\hline
 & \multicolumn{3}{c}{calculation} & & \multicolumn{3}{c}{experiment} \\ \cline{2-4} \cline{6-8}
atom & \multicolumn{3}{c}{x \ \ \ \ \ \ y \ \ \ \ \ \ z} & & \multicolumn{3}{c}{x
\ \ \ \ \ \ y \ \ \ \ \ \ z} \\ \hline
Na & \multicolumn{3}{c}{0.0392 \ \ 1/4\ \ \ 0.9910} &  &\multicolumn{3}{c}{
0.0328 \ \ 1/4\ \ -0.0065} \\
O1 & \multicolumn{3}{c}{0.4919 \ \ 1/4\ \ \ 0.0885} & &\multicolumn{3}{c}{
0.4834 \ \ 1/4\ \ \ 0.0808} \\
O2 & \multicolumn{3}{c}{0.2940 0.0428 0.7046} & &\multicolumn{3}{c}{0.2881
0.0394 0.7112} \\ \hline
\hline

\end{tabular}%
\end{table}

NaOsO$_{3}$ has an orthorhombic perovskite structure with space group of
\textit{Pnma}\cite{NaOsO3}. There are four formula units (f.u.) per unit
cell, and the 20 atoms in the unit cell can be classified as four
nonequivalent crystallographic sites: Na, Os, O1 and O2 according to the
symmetry. They are located at 4\textit{c}, 4\textit{b}, 4\textit{c} and 8%
\textit{d} sites, respectively and result in seven internal coordinates.
From the X--ray diffraction experiment\cite{NaOsO3}, the lattice constants
of NaOsO$_{3}$ are determined to be \textit{a}=5.384 \AA , \textit{b}=7.580
\AA\ and \textit{c}=5.328 \AA , respectively. Based on the experimental
lattice parameters, we optimize all independent internal atomic coordinates
until the corresponding forces are less than 1 mRy/a.u. We confirm that the
Coulomb $U$ and SOC have only small effect on the crystal structure and list
in Table I the internal atomic coordinates by LDA calculations. Our
numerical internal coordinates are in good agreement with the experimental
result, as shown in Table 1.

\begin{figure}[tbp]
\includegraphics[scale=0.4]{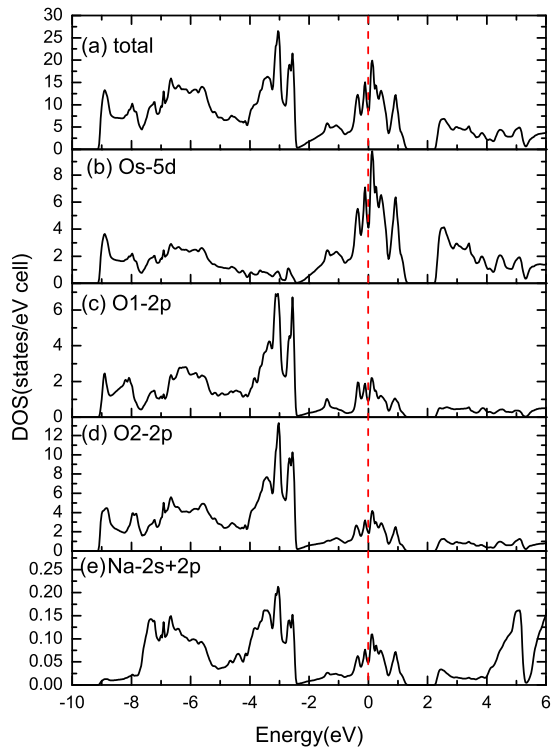}
\caption{Electronic density of states (DOS) from LDA calculation. Fermi
energy E$_{f}$ is set to zero. (a) TDOS, (b) Os 5\textit{d} PDOS, (c) O1 2%
\textit{p} PDOS, (d) O2 2\textit{p} PDOS, (e) Na PDOS.}
\end{figure}

\begin{figure}[tbp]
\includegraphics[scale=0.4]{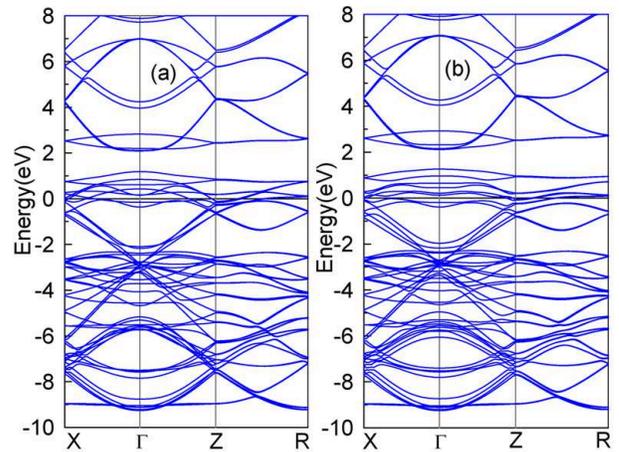}
\caption{Band structure of NaOsO$_{3}$, shown along the high symmetry
directions. (a) LDA, (b) LDA+SO.}
\end{figure}

\section{RESULTS AND DISCUSSIONS}
Using the experimental lattice constants and the numerical internal
coordinates, we first perform non--magnetic LDA calculation. The total
density of states (TDOS), Os 5\textit{d} partial density of states (PDOS),
O1 2\textit{p }PDOS, O2 2\textit{p }PDOS, Na 2\textit{s}, 2\textit{p} PDOS
has been plotted in Fig.1(a)-(e), respectively. Our TDOS is very similar to
that obtained based on the experimental crystal structure (See Fig.4a of Ref.%
\cite{NaOsO3}). The energy range, -9.0 to -2.4 eV is dominated by O1 2%
\textit{p} and O2 2\textit{p} bands with a small contribution from Os 5%
\textit{d} state. Both Na 2\textit{s} and 2\textit{p} states, appearing
mainly above 4 eV, have also considerable distribution between -9.0 to -2.4
eV, where O 2\textit{p} state is mainly located, indicating the
non--negligible hybridization between Na and O states despite that Na is
highly ionic. The Os atom is octahedrally coordinated by six O\ atoms,
making the Os 5\textit{d} band to split into the t$_{2g}$ and e$_{g}$
states, and the 12 t$_{2g}$ bands are located from -2.8 to 1.2 eV, as shown
in Fig. 2a. Due to the extended nature of 5\textit{d }states, the crystal
splitting between t$_{2g}$ and e$_{g}$ states is large, and the e$_{g}$
states are located about 2.0 eV higher than the Fermi energy (E$_{f}$) and
disperse widely. While providing the basic features of the electronic
structure, LDA produces a metallic state due to partially occupied Os 5%
\textit{d} t$_{2g}$\ band.

SOC of 5\textit{d} electrons is about 0.5 eV\cite{SOC}, which is one order
of magnitude\ larger than that of 3\textit{d} electrons. Therefore, SOC
usually changes the 5\textit{d} band dispersion significantly and plays an
essential role in the gap opening of Sr$_{2}$IrO$_{4}$ as well as of
pyrochlore iridates\cite{Sr2IrO4-1,Sr2IrO4-2,Weyl-semimetal}. In order to
investigate the effect of SOC on the electronic structure, we compare the
results obtained in the presence and absence of SOC, which are given in Fig.
2. The difference between the bands with and without SOC is small, as
demonstrated in Fig.2. For 5\textit{d}$^{5}$ electronic configuration of Sr$%
_{2}$IrO$_{4}$\cite{Sr2IrO4-1}, and A$_{2}$Ir$_{2}$O$_{7}$\ (\textit{A}=Y or
rare earth)\cite{Weyl-semimetal}, where J$_{eff}=1/2$ picture is valid, SOC
has a dramatic effect on the band structure. In NaOsO$_{3}$, Os occurs in
its 5$^{+}$ valence and there are 3 electrons in its t$_{2g}$ band. Since t$%
_{2g}$ band is half filled, it is natural to expect the effect of SOC to be
small. As shown in Fig. 1a, the Fermi level is located near a sharp peak in
the DOS. The relatively high density of states at the Fermi energy (N(E$_{f}$%
))\ suggests the possibility of a Stoner instability against ferromagnetism
(FM). Therefore, we perform a spin polarized calculation, however our
LSDA+SO calculation with initial FM setup converges to the non--magnetic
state. Thus consistent with Shi et al.\cite{NaOsO3}, the FM state is not
stable in LSDA+SO\ calculation.

\begin{figure}[tbp]
\includegraphics[scale=0.4]{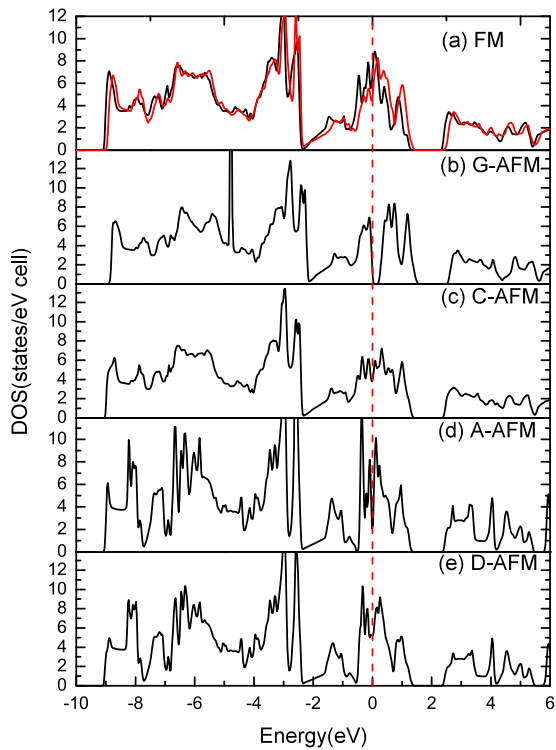}
\caption{Electronic density of states (DOS) from LDA+SO+U (U=2.0 eV)
calculation. Fermi energy E$_{f}$ is set to zero. (a) FM configuration, (b)
G-AFM (c) C-AFM, (d) A-AFM (e) D-AFM. The black/red line is for
up/down-spin, respectively. For AFM configurations, spin--up is the same as
spin--down, thus only one spin channel is plotted.}
\end{figure}

Although the 5\textit{d} orbitals are spatially extended, it has been found
that the electronic correlations are important for 5\textit{d} TMO\cite%
{Sr2IrO4-1,Sr2IrO4-2}. Moreover, the experiment reveal that NaOsO$_{3}$ has
a long-range magnetic order at low temperature\cite{NaOsO3}. We therefore
utilize LSDA+U scheme\cite{+U}, which is adequate for searching for
magnetically ordered insulating ground states \cite{LDA+DMFT}. Although, the
accurate value of \textit{U} is not known for perovskite osmates, the
estimates of the values of $U$ have been recently obtained between 1.4 and
2.4 eV in layered Sr$_{2}$IrO$_{4}$/Ba$_{2}$IrO$_{4}$ \cite{cRPA}. We
generally expect the screening to be larger in three dimensional (3D)
systems than in two dimensional (2D) ones, and one can image that \textit{U}
in NaOsO$_{3}$ should be smaller than that in Sr$_{2}$IrO$_{4}$/Ba$_{2}$IrO$%
_{4}$. We therefore perform LDA+U+SO calculation and vary parameter $U$\
between 0.5 and 2.0 eV. Numerical results show that the electronic
correlations can stabilize the FM configuration and narrow the Os t$_{2g}$
band. However, as shown in Fig.3(a), our LDA+U+SO calculation with $U$= 2.0
eV still gives a metallic solution. Naively, one may expect that using
larger Coulomb $U$ will result in an insulating state. However, consistent
with Shi et al.\cite{NaOsO3}, our additional calculations show that
increasing $U$ cannot solve this problem, and even a quite large \textit{U}
(=6.0 eV) cannot open the band gap. Therefore, electronic correlations alone
cannot explain the insulating behavior, and the MIT is not of a Mott--type.

After studying the effect of SOC and \textit{U}, we subsequently investigate
the effect of various magnetic orders. We considered four antiferromagnetic
(AFM) configurations besides the FM state: A--type AFM state (A--AFM) with
layers of Os ions coupled ferromagnetically in a given set of (001) planes
but with alternate planes having opposite spin orientation; C--type AFM
state (C--AFM) with lines of Os ions coupled ferromagnetically in a given
direction (001) but with alternate lines having opposite spin orientation;
G--type AFM state (G--AFM) with Os ions coupled antiferromagnetically with
all of their nearest neighbors; D--type AFM state (D--AFM) where Os ions
lying within alternating planes perpendicular to [001] direction are coupled
ferromagnetically along either [010] or [100] directions while different
lines are coupled antiferromagnetically. Same as with the FM setup, the
LSDA+SO calculation with \textit{U}=0 for all considered AFM setups
converges to the nonmagnetic metallic state.

\begin{table}[tbp]
\caption{Spin $\langle S\rangle $ and orbital $\langle O\rangle $\ moment
(in $\protect\mu _{B}$) as well as the total energy E$_{tot}$ per unit cell
(in eV) for several magnetic configurations, as calculated using LDA+U+SO
method with U=2.0 eV. (E$_{tot}$ is defined relative to the G--AFM
configuration.).}%
\begin{tabular}{llllll}
\hline
\hline
Configuration & G--AFM & FM & C--AFM & A--AFM & D--AFM \\ \hline
E$_{tot}$ & 0 & 0.243 & 0.186 & 0.282 & 0.205 \\
$\langle S\rangle $ & 0.94 & 0.22 & 0.54 & 0.29 & 0.20 \\
$\langle O\rangle $ & -0.11 & -0.01 & -0.04 & -0.03 & -0.03 \\ \hline
\hline
\end{tabular}%
\end{table}

On the other hand, the non--zero Coulomb interaction \textit{U }of Os 5%
\textit{d} is found to stabilize the AFM configuration. Our calculation
confirms that the magnetic order has a significant effect, and for a
reasonable \textit{U} ($\leq $2.0 eV), G--AFM configuration\ is the only
insulating solution as shown in Fig.3. Moreover, regardless the value of
\textit{U}, the G--AFM configuration always has the lowest total energy.
Thus we believe G--AFM configuration is the magnetic ordering state observed
by the experiment \cite{NaOsO3} . With increasing \textit{U} the band
structure will change, but only when \textit{U} is larger than 1.0 eV, the
G--AFM solution becomes insulating. The DOS from U=1.0 eV (see the Fig.4c of
Ref.21) is similar with that from U=2.0 eV (see Fig.3b of present work),
which again indicates that the Coulomb U is not efficient to open the band
gap. It is found that the magnetic moment is mainly located at Os site, and
despite of strong hybridization between Os 5\textit{d }and O\textit{\ 2p}, O
site is basically non--magnetic (less than 0.003 $\mu _{B}$). The numerical
data for \textit{U}=2.0 eV are given in Table II. For the 5\textit{d}$^{5}$
electronic systems such as BaIrO$_{3}$, Sr$_{2}$IrO$_{4}$, pyrochlore
iridates etc, it has been found that due to the strong spin--orbit
entanglement in 5\textit{d} states, the magnetic orbital moment is about
twice larger than the spin moment\cite%
{Sr2IrO4-1,Sr2IrO4-2,Weyl-semimetal,Orb moment in BaIrO3}, even in the
presence of strong crystal field and band effects. Contrary to 5\textit{d}$%
^{5}$ systems, the obtained orbital moment for NaOsO$_{3}$ is much smaller
than its spin moment, showing again that SOC effect is small for this 5%
\textit{d}$^{3}$ electronic configuration case. As shown in Table II, the
magnitude of magnetic moment is sensitive to the magnetic configuration,
indicating the itinerant nature of magnetism. For the same \textit{U} value,
the G--AFM\ configuration always has the largest magnetic moment among the
considered states. However, as shown in Table II, our numerical magnetic
moment (0.83 $\mu _{B}$) is much smaller than the experimental one\cite%
{NaOsO3}. For an itinerant magnet, one may still fit the $\chi (T)$\ curve
by the Curie--Weiss law, but cannot estimate the magnetic moment accurately
based on the Curie--Weiss constant\cite{Moriya book}. Thus, the experimental
magnetic moment may not be reliable. The energy difference between various
magnetic configurations is large, which is consistent with the observed high
magnetic transition temperature (about 410 K)\cite{NaOsO3} although here one
cannot estimate the interatomic exchange interaction and T$_{N}$ based on
the difference between total energies accurately as in the local moment
systems\cite{Wan}. Since the G--AFM configuration is the only insulating
state, it is easy to understand that both magnetic and electronic phase
transitions occur at the same temperature and our calculation indeed
confirms that the MIT\ of NaOsO$_{3}$ is a Slater--type transition.

\section{SUMMARY}
In summary, we have investigated the detailed electronic structure and
magnetic properties of NaOsO$_{3}$ using full potential linearized augmented
plane wave method. Our results show that the electronic structure near the
Fermi energy E$_{f}$ is dominated by strongly hybridized Os 5\textit{d} and
O 2\textit{p} states. Despite of its big value the SOC has only weak effect
on the band structure and magnetic moment. The electronic correlations alone
cannot open the band gap, and the low temperature phase of NaOsO$_{3}$ is
not a Mott--type insulator. The magnetic configuration has an important
effect on the conductivity, and the ground state is a G--type AFM insulator.
It is the interplay of the Coulomb interaction and magnetic ordering that
result in the insulating behavior of NaOsO$_{3}$.

\section{ACKNOWLEDGMENTS}
The work was supported by the National Key Project for Basic Research of
China (Grant no. 2011CB922101 and 2010CB923404), NSFC under Grant no.
91122035, 11174124 and 10974082. The project also funded by Priority
Academic Program Development of Jiangsu Higher Education Institutions. S.Y.S
was supported by DOE Computational Material Science Network (CMSN) and DOE
SciDAC Grant No. SE-FC02-06ER25793



\begin{thebibliography}{99}

\bibitem{Review M-I} M. Imada, A. Fujimori, and Y. Tokura, Rev. Mod. Phys.
\textbf{70}, 1039 (1998).

\bibitem{CMR} S. Jin, T. H. Tiefel, M. McCormack, R. A. Fastnacht, R. Ramesh
and L. H. Chen, Science \textbf{264}, 413 (1994); P. Schiffer, A. P.
Ramirez, W. Bao, and S-W. Cheong, Phys. Rev. Lett. \textbf{75}, 3336 (1995).

\bibitem{Pickett-HTC} W. Pickett, Rev. Mod. Phys. \textbf{61}, 433 (1989).

\bibitem{Sr2IrO4-1} B.J. Kim, Hosub Jin, S. J. Moon, J.-Y. Kim, B.-G. Park,
C. S. Leem, Jaejun Yu, T. W. Noh, C. Kim, S.-J. Oh, J.-H. Park, V. Durairaj,
G. Cao, and E. Rotenberg, Phys. Rev. Lett. \textbf{101}, 076402 (2008); B.J.
Kim, H. Ohsumi, T. Komesu, S. Sakai, T. Morita, H. Takagi, and T. Arima,
Science \textbf{323}, 1329 (2009).

\bibitem{Sr2IrO4-2} H. Jin, H. Jeong, T. Ozaki and J. Yu, Phys. Rev. B
\textbf{80}, 075112 (2009).

\bibitem{SOC} L.F. Mattheiss, Phys. Rev. B \textbf{13}, 2433 (1976).

\bibitem{GME} S. Chikara, O. Korneta, W. P. Crummett, L. E. DeLong, P.
Schlottmann and G. Cao, Phys. Rev. B \textbf{80}, 140407 (R) (2009).

\bibitem{HTC} F. Wang and T. Senthil, Phys. Rev. Lett. \textbf{106}, 136402
(2011).

\bibitem{Weyl-semimetal} X. Wan, A.M. Turner, A. Vishwanath, and S.Y.
Savrasov, Phys. Rev. B \textbf{83}, 205101 (2011).

\bibitem{Axion insulator} X. Wan, A. Vishwanath, and S. Y. Savrasov,
Phys. Rev. Lett \textbf{108}, 146601 (2012).

\bibitem{Pesin-Balents} D.A. Pesin and L. Balents, Nature Physics \textbf{6}%
, 376 (2010).

\bibitem{Kim} H.-M. Guo and M. Franz, Phys. Rev. Lett \textbf{103}, 206805
(2009). B. J. Yang, Y. B. Kim, Phys. Rev. B \textbf{82}, 085111 (2010). M.
Kargarian, J. Wen, G. A. Fiete, Phys. Rev. B \textbf{83}, 165112 (2011).

\bibitem{Bi2Ir2O7} T. F. Qi, O. B. Korneta, X. Wan, G. Cao, arXiv:1201.0538
(2012).

\bibitem{Kitav} G. Jackeli and G. Khaliulin, Phys. Rev. Lett. \textbf{102},
017205 (2009).


\bibitem{Cd2Os2O7-1} W. Sleight, J. L. Gillson, J. F. Weiher, and W.
Bindloss, Solid State Commun. \textbf{14}, 357 (1974).

\bibitem{Cd2Os2O7} D. Mandrus, J.R. Thompson, R. Gaal, L. Forro, J.C. Bryan,
B.C. Chakoumakos, L.M. Woods, B.C. Sales, R.S. Fishman, and V. Keppens,
Phys. Rev. B \textbf{63}, 195104 (2001); W. J. Padilla, D. Mandrus and D. N.
Basov, Phys. Rev. B \textbf{66}, 035120 (2002).

\bibitem{Cd2Os2O7-3} Y. H. Matsuda, J. L. Her, S. Michimura, T. Inami, M.
Suzuki, N. Kawamura, M. Mizumaki, K. Kindo, J. Yamauara, and Z. Hiroi, Phys.
Rev. B \textbf{84}, 174431 (2011).

\bibitem{Cd2Os2O7-4} A. Koda, R. Kadono, K. Ohishi, S. R. Saha, W. Higemoto,
S. Yonezawa, Y. Muraoka, Z. Hiroi, J. Phys. Soc. Japan \textbf{76}, 063703
(2007).

\bibitem{Singh LSDA} D. J. Singh, P. Blaha, K. Schwarz and J. O. Sofo, Phys.
Rev. B \textbf{65}, 155109 (2002); H. Harima, J. Phys. Chem. Solids \textbf{%
63}, 1035 (2002).

\bibitem{Cd2Os2O7-5} H. Shinaoka, T. Miyake, S. Ishibashi, arXiv:1111.6347
(2011).

\bibitem{NaOsO3} Y.G. Shi, Y.F. Guo, S. Yu, M. Arai, A.A. Belik, A. Sato, K.
Yamaura, E. Takayama-Muromachi, H.F. Tian, H.X. Yang, J.Q. Li, T. Varga,
J.F. Mitchell, and S. Okamoto, Phys. Rev. B \textbf{80}, 161104 (2009).

\bibitem{BaNiOsO} A. S. Erickson, S. Misra, G. J. Miller, R. R. Gupta, Z.
Schlesinger, W. A. Harrison, J. M. Kim, and I. R. Fisher, Phys. Rev. Lett.
\textbf{99}, 016404 (2007).

\bibitem{Pickett} K.-W. Lee, W. E. Pickett, EPL \textbf{80}, 37008 (2007).

\bibitem{RbOs2O6} Z. Hiroi, S. Yonezawa, Y. Muraoka, J. Phys. Soc. Jpn.
\textbf{73}, 1651 (2004); , R. Saniz, J. E. Medvedeva, L.-H. Ye, T.
Shishidou, and A. J. Freeman, Phys. Rev. B \textbf{70}, 100505 (2004).

\bibitem{KOs2O6 DFT} J. Kune\v{s}, T. Jeong and W.E. Pickett, Phys. Rev. B
\textbf{70}, 174510 (2004).

\bibitem{Slater state} J. C. Slater, Phys. Rev. \textbf{82}, 538 (1951).

\bibitem{WIEN} P. Blaha, K. Schwarz, G. K. H. Madsen, D. Kvasnicka, and J.
Luitz, WIEN2k, An Augmented Plane Wave + Local Orbitals Program for
Calculating Crystal Properties (Karlheinz Schwarz, Technische Universit\"{a}%
t Wien, Austria), 2001, ISBN3--9501031--1--2.

\bibitem{LSDA-good-1} K. Maiti, Solid State Commun. \textbf{149}, 1351
(2009); K. Maiti, Phys. Rev. B \textbf{73}, 115119 (2006).

\bibitem{LSDA-good-2} D. J. Singh, J. Appl. Phys. \textbf{79}, 4818 (1996);
A. T. Zayak, X. Huang, J. B. Neaton, and K. M. Rabe, Phys. Rev. B \textbf{77}%
, 214410 (2008); X. Wan, J. Zhou, and J. Dong, Europhys. Lett. \textbf{92},
57007 (2010).

\bibitem{SOC in WIEN2k} D.D. Koelling, B.N. Harmon, J. Phys. C \textbf{10},
3107 (1977).

\bibitem{+U} V.I. Anisimov, F. Aryasetiawan, and A.I. Lichtenstein, J.
Phys.: Condens. Matter \textbf{9}, 767 (1997).

\bibitem{LDA+DMFT} G. Kotliar, S. Y. Savrasov, K. Haule and V. S.
Oudovenko,\ Rev. Mod. Phys. \textbf{78}, 865 (2006).

\bibitem{cRPA} R. Arita, J. Kune\v{s}, A.V. Kozhevnikov, A.G. Eguiluz, M.
Imada, arXiv:1107.0835 (2011).

\bibitem{Orb moment in BaIrO3} M. A. Laguna-Marco1, D. Haskel, N.
Souza-Neto, J. C. Lang, V. V. Krishnamurthy, S. Chikara, G. Cao, and M. van
Veenendaal, Phys. Rev. Lett. \textbf{105}, 216407 (2010).

\bibitem{Moriya book} T. Moriya, \textit{Spin fluctuations in itinerant
electron magnetism}, (Springer-Verlag, 1985).

\bibitem{Wan} X. Wan, M. Kohno and X. Hu, Phys. Rev. Lett. \textbf{94},
087205 (2005). %
\end{thebibliography}
\end{document}